\documentclass[twocolumn,aps,prb,groupedaddress,superscriptaddress,amsmath,floatfix,amssymb,showpacs,noeprint,english]{revtex4-2}
\usepackage{graphicx}
\usepackage{dcolumn}
\usepackage{bm}
\usepackage[T1]{fontenc}
\usepackage{mathrsfs}
\usepackage{amsbsy}
\usepackage{amstext}
\usepackage{amsmath, amsthm, amsfonts, amssymb, bbold}
\usepackage{setspace}
\usepackage{esint}
\usepackage{xcolor,colortbl}
\usepackage{float}
\usepackage{placeins}
\usepackage{algorithm}
\usepackage{algpseudocode}
\usepackage[colorlinks,citecolor=blue,urlcolor=blue,linkcolor=blue,hypertexnames=true]{hyperref} 
\usepackage{babel}
\usepackage{booktabs}
\usepackage{tikz}

\usepackage{tikz}
\usepackage{quantikz}
\usepackage{threeparttable}
\usepackage{multirow}
\usepackage{float}

\begin{document}

\title{Warm-Starting QAOA with XY Mixers: A Novel Approach for\\ Quantum-Enhanced Vehicle Routing Optimization}

\author{Rafael~Simões~do~Carmo}
\affiliation{Faculty of Sciences, UNESP - S{\~a}o Paulo State University, 17033-360 Bauru-SP, Brazil}

\author{Marcos~C.~S.~Santana}
\affiliation{Faculty of Sciences, UNESP - S{\~a}o Paulo State University, 17033-360 Bauru-SP, Brazil}

\author{F. F. Fanchini}
\affiliation{Hospital Israelita Albert Einstein, 05652-900 São Paulo-SP, Brazil}
\affiliation{Faculty of Sciences, UNESP - S{\~a}o Paulo State University, 17033-360 Bauru-SP, Brazil}
\affiliation{QuaTI - Quantum Technology \& Information, 13560-161 São Carlos-SP, Brazil}

\author{Victor~Hugo~C.~de~Albuquerque}
\affiliation{Department of Teleinformatics Engineering, UFC - Federal University of Ceará, 60020 Fortaleza-CE, Brazil}

\author{João~Paulo~Papa}
\affiliation{Faculty of Sciences, UNESP - S{\~a}o Paulo State University, 17033-360 Bauru-SP, Brazil}

\date{\today}

\begin{abstract}
Quantum optimization algorithms, such as the Quantum Approximate Optimization Algorithm (QAOA), are emerging as promising heuristics for solving complex combinatorial problems. To improve performance, several extensions to the standard QAOA framework have been proposed in recent years. Two notable directions include: warm-starting techniques, which incorporate classical approximate solutions to guide the quantum evolution, and custom mixer Hamiltonians, such as XY mixers, which constrain the search to feasible subspaces aligned with the structure of the problem. In this work, we propose an approach that integrates these two strategies: a warm-start initialization with an XY mixer ansatz, enabling constraint-preserving quantum evolution biased toward high-quality classical solutions. The method begins by reformulating the combinatorial problem as a MaxCut instance, solved approximately using the Goemans-Williamson algorithm. The resulting binary solution is relaxed and used to construct a biased superposition over valid one-hot quantum states, maintaining compatibility with the XY mixer’s constraints. We evaluate the approach on 5-city instances of the Traveling Salesperson Problem (TSP), a canonical optimization problem frequently encountered as a subroutine in real-world Vehicle Routing Problems (VRPs). Our method is benchmarked against both the standard XY-mixer QAOA and a warm-start-only variant based on MaxCut relaxation. Results show that the proposed combination consistently outperforms both baselines in terms of the percentage and rank of optimal solutions, demonstrating the effectiveness of combining structured initializations with constraint-aware quantum evolution for optimization problems.
\end{abstract}

\maketitle
\section{Introduction}
%
%
%
%

In Intelligent Transportation Systems (ITS), efficient route planning plays a critical role in enabling real-time decision-making for logistics, delivery networks, ride-sharing services, and autonomous vehicle fleets. One of the foundational problems in this domain is the Vehicle Routing Problem (VRP) \cite{toth2002vehicle}, which seeks to determine optimal routes for a set of vehicles servicing multiple locations while satisfying constraints such as vehicle capacity, service windows, or geographic clustering. Among its many variants, the Capacitated Vehicle Routing Problem \cite{CVRP} is of particular interest due to its applicability in modern urban logistics and smart mobility services.

As classical solvers for these problems often rely on heuristics or approximations to meet runtime requirements, there remains a need for new paradigms that offer improved performance or reduced time-to-solution. In this context, quantum computing has emerged as a promising technology for tackling combinatorial optimization problems \cite{abbas2023quantum}. Especially hybrid quantum-classical algorithms such as the Quantum Approximate Optimization Algorithm \cite{farhi2014quantum} leverage quantum circuits to explore solution spaces in fundamentally different ways than their classical counterparts, with the potential to provide high-quality solutions faster. Furthermore, an extension of the original QAOA, known as the Quantum Alternating Operator Ansatz ~\cite{hadfield2019quantum, wurtz2021constraints}, provides a generalized framework for addressing optimization problems with hard constraints. This is achieved by designing custom initial states and mixers that restrict the quantum evolution to a structured subspace where some or all constraints are satisfied. While this subspace may still include infeasible solutions, it often significantly reduces the search space and guides the optimization toward higher-quality and potentially valid solutions, thereby improving the algorithm’s performance.

A well-established strategy for solving VRPs involves decomposing the problem into smaller subproblems using a cluster-first, route-second heuristic \cite{CVRP}. In this approach, customer locations are first clustered into feasible groups according to problem constraints, and then each cluster is solved as a standalone Traveling Salesperson Problem \cite{karp1972reducibility, lawler1985traveling}. The TSP, being NP-hard, poses significant computational challenges, particularly when solved repeatedly within larger ITS workflows. Feld et al.~\cite{feld2019hybrid} introduced a hybrid quantum-classical strategy for this setting, where clustering is performed classically and each TSP subproblem is addressed using quantum optimization techniques.

Motivated by this line of research \cite{herzog2024improving}, we explore a quantum approach for solving such constrained TSP subproblems more effectively. Specifically, we propose a variant of QAOA that combines warm-start initialization with a constraint-preserving XY mixer. This hybrid formulation restricts the quantum evolution to a subspace in which each tour position is assigned to a single city, thus eliminating one class of constraint violations, while using classical guidance to bias the search toward high-quality solutions. We evaluate this method on small-scale TSP instances derived from CVRP-style decompositions, with the goal of identifying strategies that can scale alongside future advances in quantum hardware.

This paper is organized as follows. Section~\ref{sec:tsp_problem} provides the necessary background on the CVRP and TSP. Section~\ref{sec:qaoa}  introduces the QAOA versions that we studied: constraint-preserving mixers such as the XY ansatz and warm-starting. In Section~\ref{sec:method}, we introduce our proposed approach, which integrates warm-start initialization with an XY mixer to guide quantum optimization within a structured subspace. Section~\ref{sec:experiments} describes the experimental setup and presents the results of numerical simulations on 5-city TSP instances. Section~\ref{sec:analysis} analyzes the impact of our method in comparison to standard baselines and discusses its applicability to real-world ITS settings. Finally, Section~\ref{sec:conclusion} concludes the work and outlines potential directions for future research.

\section{Vehicle Routing Problems and TSP}
\label{sec:tsp_problem}
\subsection{Capacitated Vehicle Routing Problem}

The Capacitated Vehicle Routing Problem is a generalization of the Traveling Salesperson Problem that plays a central role in supply chain logistics, distribution systems, and intelligent transportation. It is defined on a complete graph $G = (V, E)$, where node $0$ represents a depot and each node $i \geq 1$ represents a customer with demand $\delta_i$. Given a fleet of identical vehicles, each with capacity $C$, the objective is to determine a set of vehicle routes such that:
(i) each customer is visited exactly once,
(ii) the total demand on any route does not exceed $C$, and
(iii) the total distance (or cost) of all routes is minimized, with each route starting and ending at the depot.

Formally, the CVRP seeks a partition of the customer set into feasible subsets (routes), each forming a tour beginning and ending at the depot, with the constraint $\sum_{i \in R} \delta_i \leq C$ for every route $R$. The CVRP is NP-hard, and exact methods become intractable for large instances, making it a natural candidate for approximate and heuristic approaches.

Several recent works have explored how to encode the CVRP into the Quadratic Unconstrained Binary Optimization (QUBO) formalism, which allows the application of quantum and quantum-inspired optimization algorithms. In particular, Feld et al.~\cite{feld2019hybrid} and related efforts~\cite{vyskocil2019embedding} propose encodings that jointly capture routing and capacity constraints, albeit at the cost of requiring a large number of binary variables (and thus qubits). Even for modest-sized instances, these formulations demand hundreds of qubits, which exceed the capabilities of current quantum hardware.

To mitigate this, problem decomposition strategies have been proposed, wherein the CVRP is reduced to a sequence of smaller subproblems that can be tackled individually. One widely adopted decomposition is the \textit{cluster-first, route-second} heuristic, where customers are first grouped into capacity-feasible clusters, and each cluster is then solved as an independent TSP. This motivates our focus on the TSP as a core quantum subroutine with potential relevance to practical CVRP solvers.

\subsection{Traveling Salesperson Problem}

The Traveling Salesperson Problem is one of the most well-known NP-hard problems in combinatorial optimization~\cite{karp1972reducibility}. It is defined as follows: given a complete, weighted graph $ G = (V, E) $, where each edge $ (u, v) \in E $ has an associated cost $ W_{uv} $, the goal is to find the shortest possible Hamiltonian cycle. That is, a closed path that visits each vertex in $ V $ exactly once and returns to the starting vertex, with no repeated nodes or subcycles.

In this work, we consider the standard symmetric version of the problem, where the cost matrix is symmetric (i.e., $ W_{uv} = W_{vu} $) and the graph is fully connected. Despite its deceptively simple formulation, the TSP remains computationally intractable for large instances and continues to serve as a benchmark for evaluating both classical and quantum optimization algorithms~\cite{twostep, grovertsp}.

\subsection{QUBO Formulation of TSP}

We adopt the QUBO formulation of the TSP proposed by Lucas~\cite{lucas2014ising}. To reduce the number of binary variables, we fix the starting node of the tour without loss of generality. This reduces the number of binary variables from $ N^2 $ to $ (N - 1)^2 $, a simplification that is particularly relevant for small instances and current quantum hardware with limited qubit counts.

Let $ x_{i,t} \in \{0,1\} $ be a binary variable that is 1 if city $ i $ is visited at position $ t $ in the tour, and 0 otherwise. Here, $ i \in \{1, \dots, N-1\} $ and $ t \in \{1, \dots, N-1\} $, assuming city 0 is fixed at position 0 in the cycle.

The QUBO formulation includes two main constraints:

\begin{enumerate}
    \item \textbf{Each city must be visited exactly once:}
    \begin{equation}
        \sum_{t=1}^{N-1} x_{i,t} = 1 \quad \forall i \in \{1, \dots, N-1\}
    \end{equation}
    
    \item \textbf{Each position in the tour must be occupied by exactly one city:}
    \begin{equation}
        \sum_{i=1}^{N-1} x_{i,t} = 1 \quad \forall t \in \{1, \dots, N-1\}
    \end{equation}
\end{enumerate}

These constraints are incorporated into the QUBO objective function using penalty terms. The complete QUBO cost function includes both the distance minimization objective and the penalty terms for constraint violations:

\begin{equation}
    H_{\text{QUBO}} = H_{\text{cost}} + A \cdot H_{\text{row}} + B \cdot H_{\text{col}},
\end{equation}

where:
- $ H_{\text{cost}} $ encodes the total distance of the tour,
- $ H_{\text{row}} $ penalizes cities appearing more than once,
- $ H_{\text{col}} $ penalizes tour positions being occupied by more than one city,
- $ A $ and $ B $ are positive penalty coefficients chosen to enforce constraint satisfaction.

The cost term is defined as:
\begin{equation}
    H_{\text{cost}} = \sum_{i,j=1}^{N-1} \sum_{t=1}^{N-2} W_{ij} \cdot x_{i,t} x_{j,t+1} + \sum_{i=1}^{N-1} W_{0i} \left(x_{i,1} + x_{i,N-1} \right),
\end{equation}
where $ W_{ij} $ is the distance between cities $ i $ and $ j $, and city 0 is fixed as the start and end point of the tour.

This QUBO encoding enables the application of quantum optimization algorithms, such as QAOA, to the TSP by mapping the problem onto a cost Hamiltonian defined over binary variables. 

\section{Quantum Approximate Optimization Algorithm}
\label{sec:qaoa}
\subsection{Standard formulation}

The QAOA~\cite{farhi2014quantum} is a variational quantum algorithm designed to approximately solve combinatorial optimization problems. It encodes the cost function into a problem-specific Hamiltonian $ H_C $, whose ground state corresponds to the optimal solution, and constructs a parameterized quantum circuit that alternates between the application of $ H_C $ and a mixing Hamiltonian $ H_B $. Each binary variable $ x_i \in \{0, 1\} $ is mapped to a qubit using the relation $ x_i = (1 - \hat{Z}_i)/2 $, where $ \hat{Z}_i $ is the Pauli-$Z$ operator acting on qubit $i$. This encoding transforms the cost function into a diagonal operator acting on computational basis states, enabling energy-based optimization over bitstring configurations. The circuit parameters are optimized through a classical feedback loop that iteratively updates them to minimize the cost function. This hybrid quantum-classical structure makes QAOA particularly well-suited for near-term quantum devices, as the classical optimizer can partially compensate for noise and hardware imperfections, especially in shallow circuits, by adapting the parameters in response to noisy quantum measurements.

In its standard formulation, QAOA starts from the initial state $ \ket{-}^{\otimes n} $, which is an eigenstate of the mixer Hamiltonian 
\begin{equation}
    H_B = \sum_{j=1}^{n} \sigma_j^x,
\end{equation}
where $ \sigma_j^x $ is the Pauli-$X$ operator on qubit $j$. The QAOA ansatz applies $p$ layers of alternating unitaries generated by the cost and mixer Hamiltonians:

\begin{equation}
    \ket{\vec{\gamma}\vec{\beta}} = U(H_B, \beta_p) U(H_C, \gamma_p) \cdots U(H_B, \beta_1) U(H_C, \gamma_1) \ket{-}^{\otimes n},
\end{equation}

where $ \vec{\gamma} = (\gamma_1, \dots, \gamma_p) $ and $ \vec{\beta} = (\beta_1, \dots, \beta_p) $ are variational parameters optimized via a classical outer loop. The performance of QAOA depends critically on finding optimal parameter values, which minimize the expectation value of the cost Hamiltonian:

\begin{equation}
    \min_{\vec{\gamma}, \vec{\beta}} \bra{\vec{\gamma}, \vec{\beta}} H_C \ket{\vec{\gamma}, \vec{\beta}}.
\end{equation}

Despite its advantages, QAOA still faces significant challenges, particularly in the classical parameter optimization step. Issues such as barren plateaus, local minima, and gradient vanishing can hinder convergence as the circuit depth increases~\cite{larocca2024review, blekos2024review}. Furthermore, in the current Noisy Intermediate-Scale Quantum (NISQ) era \cite{preskill2018quantum}, hardware imperfections such as gate noise, readout errors, and decoherence can significantly degrade the quality of quantum states, especially in deeper circuits. These limitations have motivated the development of problem-specific ansätze and enhanced strategies, such as constraint-preserving mixers and warm-started initializations, that reduce the effective search space and improve convergence toward high-quality solutions within the limited coherence and gate fidelity available on NISQ devices.

\subsection{Constraint-Preserving XY Mixer}

One of the most prominent extensions of the original QAOA framework is the Quantum Alternating Operator Ansatz ~\cite{hadfield2019quantum}, which introduces the ability to design custom mixer Hamiltonians and initial states that respect the structural constraints of combinatorial optimization problems. By tailoring the quantum evolution to preserve problem-specific feasibility conditions, this approach enables a more efficient and focused exploration of the solution space. As a result, it can restrict the dynamics to syntactically valid (or nearly valid) bitstrings, improving both the quality of sampled solutions and the convergence of the optimization process.

In the context of the TSP, this framework naturally leads to the design of ansätze that operate within a subspace respecting problem constraints. One of the core constraints in standard TSP encodings is that each tour position must be assigned to exactly one city, i.e., one-hot encoding within each group of qubits. To enforce this structure, we employ a block-wise initialization using $ W $-states, which are uniform superpositions over all one-hot basis states in a given register.

For example, the three-qubit $ W $-state is defined as:
\begin{equation}
    \ket{W_3} = \frac{1}{\sqrt{3}} \left( \ket{100} + \ket{010} + \ket{001} \right),
\end{equation}
where the Hamming weight is preserved. In the TSP setting, where each city position is encoded with a one-hot register of size $n$, the full initialization state is constructed as a tensor product of $W_n$ blocks:
\begin{equation}
    \ket{s_W^n} = \left( \ket{W_n} \right)^{\otimes n}.
\end{equation}

This initialization ensures that the temporal constraint (one city per position) is satisfied from the beginning of the evolution, and the quantum evolution remains within the corresponding Hamming-weight-1 subspace under the action of an appropriate mixer.

To maintain the state within this structured subspace, we adopt the XY mixer Hamiltonian:
\begin{equation}
    H_{M_{XY}} = \sum_{i=0}^{n-1} \sum_{t=0}^{n-1} \text{XY}_{(i,t),(i,t+1)},
\end{equation}
where the XY operator between qubits $ (i,t) $ and $ (j,s) $ is implemented via:
\begin{equation}
    \text{XY}_{(i,t),(j,s)} = X_{i,t} X_{j,s} + Y_{i,t} Y_{j,s},
\end{equation}
with $X$ and $Y$ denoting the Pauli-$X$ and Pauli-$Y$ matrices. These gates preserve the total excitation number (Hamming weight), allowing the system to evolve within the subspace of valid one-hot encodings per position.

Compared to the traditional X mixer used in standard QAOA, the XY mixer is more expensive to implement due to its two-qubit gate overhead, but it enforces structure in the evolution that is essential for solving constrained problems such as the TSP.

\subsection{Warm-starting QAOA}

Warm-starting QAOA (WS-QAOA) is an enhanced variant of the standard QAOA that leverages classical continuous relaxations, such as Quadratic Programming (QP) or Semidefinite Programming, to construct a more informed quantum initialization. In this approach, the classical approximation guides the quantum circuit by biasing the initial quantum state toward promising regions of the solution space, thereby improving convergence and helping mitigate issues such as barren plateaus and local minima \cite{warmstart_qaoa}. 

The quantum state initialization in WS-QAOA is constructed as follows: given a relaxed solution vector $c^* = (c_0^*, \dots, c_{n-1}^*)$ obtained from the classical preprocessing, each qubit is initialized via a single-qubit $Y$-rotation:
\begin{equation}
    |\phi^*\rangle = \bigotimes_{i=0}^{n-1} \hat{R}_Y(\theta_i) |0\rangle,
\end{equation}
where the rotation angle is defined by $\theta_i = 2\arcsin(\sqrt{c_i^*})$. This yields a biased quantum superposition that incorporates classical information while remaining compatible with variational quantum optimization.

To preserve this initialization structure during quantum evolution, WS-QAOA employs a custom mixer Hamiltonian in place of the standard transverse-field mixer $H_B = \sum_i \hat{X}_i$. The new mixer comprises single-qubit Hamiltonians defined as:
\begin{equation}
    \hat{H}^{(ws)}_{M,i} = 
    \begin{pmatrix}
    2c_i^* - 1 & -2\sqrt{c_i^*(1-c_i^*)} \\
    -2\sqrt{c_i^*(1-c_i^*)} & -2c_i^* + 1
    \end{pmatrix},
\end{equation}
which has $|\phi^*\rangle$ as its ground state. This operator can be implemented using a sequence of single-qubit rotations:
\begin{equation}
    \hat{R}_Y(\theta_i)\, \hat{R}_Z(-2\beta)\, \hat{R}_Y(-\theta_i)\, |0\rangle.
\end{equation}

However, as emphasized in the original WS-QAOA work~\cite{warmstart_qaoa}, the goal is not only to preserve the classical warm-start initialization but also to allow the quantum evolution to deviate from it in search of improved solutions. To enable this flexibility, the authors proposed a slightly modified version of the mixer, referred to as the **modified warm-start mixer**, in which the rotation elements are sign-inverted. This is the version adopted in our work:
\begin{equation}
    \hat{R}_Y(-\theta_i)\, \hat{R}_Z(-2\beta)\, \hat{R}_Y(2\theta_i)\, |0\rangle.
\end{equation}

To prevent qubits from being initialized into computational basis states $|0\rangle$ or $|1\rangle$, which would result in frozen dynamics under cost Hamiltonians diagonal in the $Z$ basis, a regularization parameter $\epsilon \in [0, 0.5]$ is introduced to avoid extreme values of $c_i^*$. The angle $\theta_i$ is computed as:
\begin{equation}
    \theta_i = 
    \begin{cases}
    2\arcsin(\sqrt{c_i^*}) & \text{if } c_i^* \in [\epsilon, 1 - \epsilon], \\
    2\arcsin(\sqrt{\epsilon}) & \text{if } c_i^* < \epsilon, \\
    2\arcsin(\sqrt{1 - \epsilon}) & \text{if } c_i^* > 1 - \epsilon.
    \end{cases}
\end{equation}

For the specific case of the maximum cut problem (MaxCut) \cite{farhi2014quantum, warmstart_qaoa} (see the definition in Appendix ~\ref{app:maxcut}), where the formulation cannot be directly relaxed into a QP without projection onto the cone of positive semidefinite matrices, WS-QAOA uses a warm-start derived from the Goemans–Williamson (GW) algorithm \cite{goemans1995improved}. The binary GW solution is then relaxed using the parameter $\epsilon > 0$, allowing the quantum evolution to explore a broader subspace beyond the initial GW assignment.

The original WS-QAOA work~\cite{warmstart_qaoa} demonstrated that setting $\epsilon = 0.25$ leads to robust performance across a range of problem instances. Furthermore, the authors showed that the regularized WS-QAOA circuit can reproduce the GW solution and, through variational refinement, surpass it. This establishes WS-QAOA as a hybrid quantum-classical algorithm that performs at least as well as classical rounding techniques, with the potential for improvement via quantum optimization.

In this work, we apply WS-QAOA to the TSP by first mapping it to an equivalent MaxCut instance, as it is well-known that any QUBO problem can be translated to an equivalent Maxcut problem \cite{QUBO_to_Maxcut} (for details, see Appendix ~\ref{app:qubo_to_maxcut}). The relaxed GW solution is used to build both the warm-started initial state and the corresponding mixer, enabling WS-QAOA to refine the classical approximation within a structured and expressive variational space.

\section{Warm-Starting with XY Mixer Ansatz}
\label{sec:method}

We propose a hybrid QAOA approach that combines warm-starting with the XY mixer ansatz to leverage both classical guidance and constraint-preserving quantum evolution. The key idea is to initialize the QAOA circuit in a quantum state that is not only biased toward an approximate classical solution, but also confined to the feasible subspace defined by the problem constraints, in this case, one-hot valid encodings for the TSP.

Our initialization begins with an approximate solution obtained by mapping the TSP instance to a MaxCut problem and solving it classically using the Goemans–Williamson algorithm. The resulting binary string $ \mathbf{z} \in \{0,1\} $ is then relaxed using a regularization factor $\epsilon = 0.25$, as proposed in~\cite{warmstart_qaoa}. The relaxed probabilities are defined as:

\begin{equation}
p_i = 
\begin{cases}
1 - \epsilon & \text{if } z_i = 1, \\
\epsilon     & \text{if } z_i = 0.
\end{cases}
\end{equation}

Rather than applying these probabilities to each qubit independently, we use them to construct a valid quantum state in the one-hot subspace, compatible with the XY mixer. For each block of qubits corresponding to a city position in the TSP tour (i.e., a set of $n$ qubits), we prepare a superposition over the one-hot basis states weighted by the square roots of the relaxed probabilities:

\begin{equation}
\left| \psi_{\text{init}}^{(k)} \right\rangle = \frac{1}{\sqrt{Z_k}} \sum_{j=1}^{n} \sqrt{p_{j}^{(k)}} \, |e_j\rangle,
\end{equation}

where $ |e_j\rangle $ denotes the one-hot state with a 1 in position $j$, $ p_j^{(k)} $ is the relaxed probability of that bit being 1 in block $k$, and $ Z_k = \sum_j p_j^{(k)} $ ensures normalization. The full initial state is the tensor product over all blocks:

\begin{equation}
\left| \Psi_{\text{init}} \right\rangle = \bigotimes_{k=1}^{n} \left| \psi_{\text{init}}^{(k)} \right\rangle.
\end{equation}

This warm-started XY mixer approach provides a principled way to combine classical solution bias with constraint-preserving quantum evolution. The initial state respects the structure imposed by the XY mixer, which preserves Hamming weight within each block during the quantum evolution, ensuring that only one city is assigned per position. At the same time, the relaxed probabilities derived from the Goemans–Williamson solution guide the quantum optimization toward promising regions of the solution space identified by the classical approximation. The complete approach is summarized in Alg \ref{algorithm1}.

Having described the overall workflow, we are now in a position to evaluate the effectiveness of our approach through numerical simulations. We compare its performance against two baselines: the pure warm-starting strategy applied after converting the TSP to a MaxCut problem, and the pure XY mixer approach initialized with a uniform product state of $W$ states.

\begin{algorithm}[H]
\caption{Warm-Started QAOA with XY Mixer (Depth-$p$)}
\label{algorithm1}
\textbf{Input:} Complete TSP graph $G = (V, E)$ with weights $W$; number of QAOA layers $p$\\
\textbf{Output:} Optimized quantum state for TSP sampling
\begin{algorithmic}[1]
\State \textbf{Convert} TSP to equivalent MaxCut instance $(V', E')$
\State \textbf{Solve} MaxCut classically using GW algorithm
\State \textbf{Relax} GW solution using regularization factor $\epsilon = 0.25$
\State \textbf{Build} warm-started initial state using relaxed probabilities:
$$
\ket{s_{\text{warm}}} = \bigotimes_{t=1}^{n} \left( \sum_{i=1}^{n} \sqrt{a_{i,t}} \ket{e_{i,t}} \right), \quad \text{where } \sum_i a_{i,t} = 1
$$
\For{$l = 1$ to $p$}
    \State Apply cost unitary: $U_C(\gamma_l) = e^{-i \gamma_l H_C}$
    \State Apply XY mixer: $U_M(\beta_l) = e^{-i \beta_l H_{M_{XY}}}$
\EndFor
\State Measure state in the computational basis to sample bitstrings $x$
\end{algorithmic}
\end{algorithm}

\section{EXPERIMENTAL RESULTS}
\label{sec:experiments}

\subsection{Methods}

To evaluate the performance of our proposed approach, we conducted numerical simulations on 5-city instances of the TSP, requiring 16 qubits under the employed encoding scheme. Since our method integrates two strategies previously explored in the literature (warm-started QAOA after a MaxCut transformation and the use of XY mixers to constrain quantum evolution to feasible subspaces) we benchmarked our combined approach against these individual components. Specifically, we compared it to: (i) warm-started QAOA with a custom mixer following a TSP-to-MaxCut mapping, and (ii) standard QAOA with an XY mixer initialized in a uniform superposition over one-hot states, ensuring a Hamming weight of one per city block. For completeness, we also included standard QAOA with the conventional X mixer as a baseline.

In both our method and the pure warm-starting variant, the classical MaxCut solution was obtained using the GW approximation algorithm, and the binary result was relaxed using a regularization factor of $\epsilon = 0.25$, as this value was shown to yield the best performance in the original warm-start QAOA study \cite{warmstart_qaoa}. All simulations were performed in a noise-free environment using the \texttt{statevector} method of Qiskit's \texttt{AerSimulator} and the COBYLA classical optimizer. We tested 10 random instances of complete graphs, each with 5 different parameter initializations and circuit depths of up to 3 layers. Simulations were executed on a machine equipped with an NVIDIA GeForce RTX 3080 Ti GPU with 12~GB of VRAM, and each full benchmark (per algorithm variant) required approximately 3 days to complete. Our focus on shallow-depth circuits reflects the practical limitations of current NISQ devices.

In order to quantitatively compare the performance of the QAOA variants, we employed evaluation metrics that emphasize the algorithm’s ability to recover the optimal solution rather than merely approximating it. Specifically, we used:

\begin{enumerate}
    \item \textbf{Percentage of True Solutions:} The fraction of sampled bitstrings that exactly correspond to the optimal TSP tour.
    
    \item \textbf{Sampling Rank:} The rank position of the optimal solution among all sampled bitstrings, ordered by their frequency of occurrence.
\end{enumerate}

These metrics are particularly suitable for combinatorial optimization problems such as the TSP, where the objective is not just to approximate the cost, but to identify the correct discrete configuration. While approximation ratios (defined as the ratio between the cost function value of the best sampled solution and that of the true optimum) are commonly reported in quantum optimization benchmarks, they do not necessarily reflect the algorithm’s success in finding the actual solution. As highlighted in prior studies~\cite{bhattachayra2025solving}, high approximation ratios may still be associated with infeasible or suboptimal solutions. For this reason, we focused on evaluation criteria that provide a more faithful assessment of the algorithm's ability to sample the true optimum.

\onecolumngrid
\begin{figure}[ht]
    \centering
    \includegraphics[width=\textwidth]{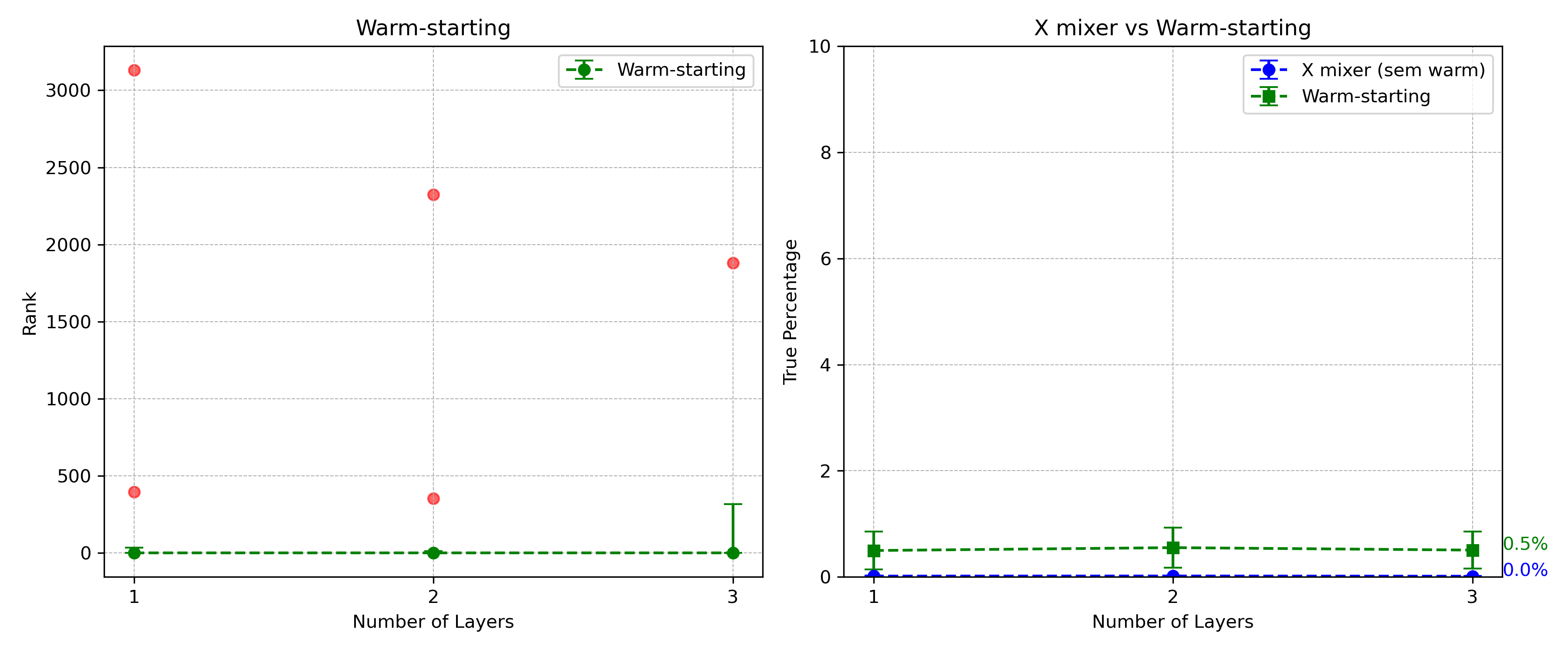} 
    \begin{minipage}{\textwidth}
    \caption{Comparison between the warm-started QAOA approach and the standard X mixer. \textbf{Left:} Sampling Rank of the optimal TSP solution across different layers for the warm-starting method. Green markers indicate the average rank, while red dots highlight outlier instances where the optimal solution was ranked far below the average. \textbf{Right:} Percentage of True Solutions recovered by each method. While warm-starting leads to marginal gains over the X mixer in terms of success rate, it frequently ranks the optimal solution as the most sampled bitstring, except for the noted outliers.}
    \label{fig:comparison_warm_vs_no_warm}
    \end{minipage}
\end{figure}
$\frac{}{}$
\onecolumngrid
\begin{figure}[h]
    \centering
    \includegraphics[width=\textwidth]{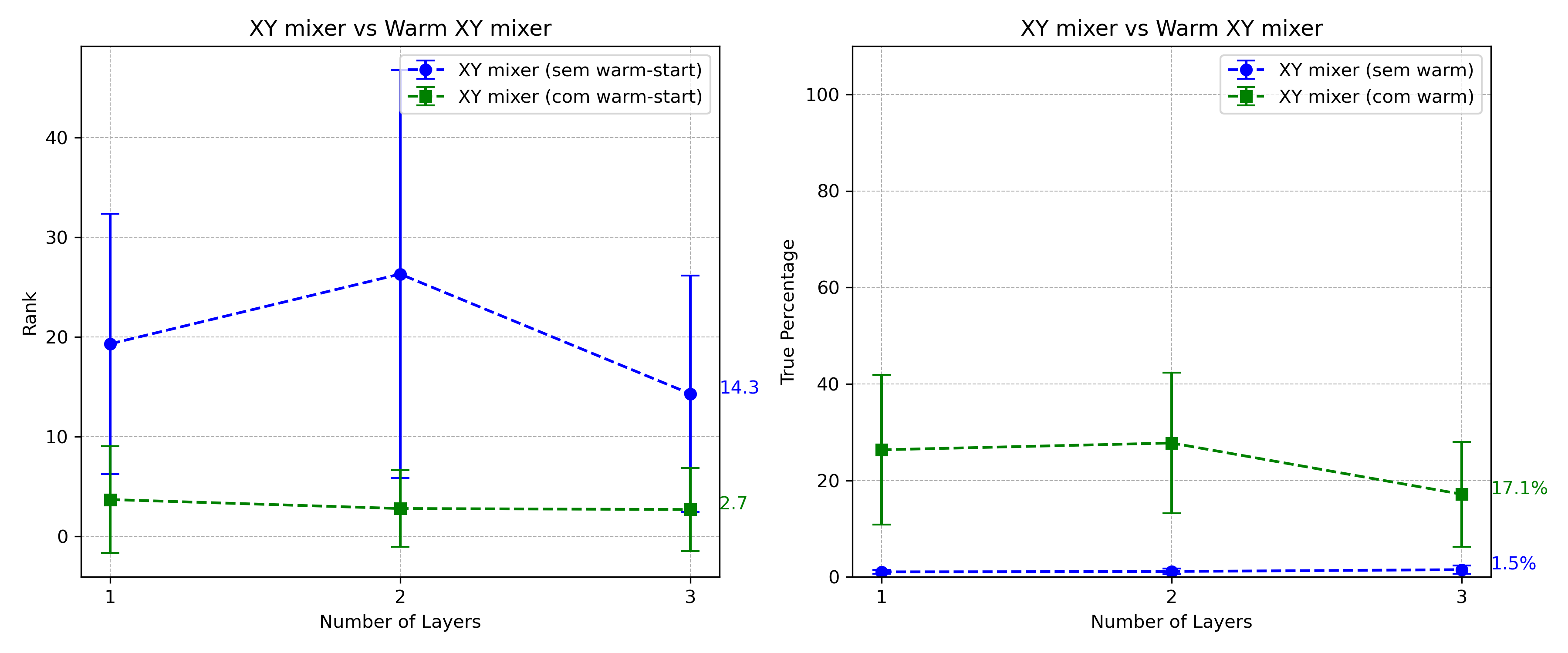} 
    \caption{Comparison between the standard XY mixer and the proposed warm-started XY mixer approach. \textbf{Left:} Sampling Rank of the optimal TSP solution across different circuit depths ($p=1,2,3$). The warm-started variant consistently ranks the optimal solution among the top 3 most frequently sampled bitstrings, while the standard XY mixer shows significantly worse and more variable ranking. \textbf{Right:} Percentage of True Solutions recovered by each method. The warm-started XY mixer achieves substantially higher success rates, peaking at $28.1\%$ for $p=2$ and maintaining strong performance across all depths. These results highlight the synergistic benefit of combining warm-start initialization with a structured mixer.}
    \label{fig:comparison_warmXY_vs_XY}
\end{figure}
\twocolumngrid

\subsection{Results and Analysis}
\label{sec:analysis}

The results in Figures~\ref{fig:comparison_warm_vs_no_warm} and \ref{fig:comparison_warmXY_vs_XY} reveal important insights into the performance of the tested QAOA variants. As expected, the standard X mixer produced the poorest results, consistently failing to sample the optimal TSP tour, or finding it only once or twice across all tested layers. While the warm-starting strategy applied to the X mixer led to modest improvements (reaching a maximum of $0.5\%$ true solutions at $p=3$) its performance remains far below that of our proposed method. Notably, the pure warm-starting approach often ranked the optimal solution as the most frequently sampled bitstring, indicating strong guidance toward high-quality regions of the solution space, despite the low overall percentage of true solutions. An exception to this behavior is observed in a few outlier cases, which will be discussed later.

Similarly, the standard XY mixer, which preserves some feasibility by operating within a fixed Hamming weight subspace, achieved better results than the X-based methods, but still lagged significantly behind the warm-started XY variant. These findings highlight that, although warm-starting and XY mixers give improvement to the standard QAOA independently, their combination leads to a substantial and synergistic performance boost.

We note that the Sampling Rank results for the standard X mixer were omitted from the analysis, as the optimal solution consistently appeared with extremely low frequency, often ranking beyond the 1000th position, so we chose to omit them.

Our proposed approach, which integrates warm-starting with an XY mixer, achieves markedly superior results. In terms of \textit{Percentage of True Solutions}, it reaches the mean percentage of $28.1\%$ at $p=2$ layers, an order of magnitude higher than any other tested variant. The \textit{Sampling Rank} results also reinforce this finding, with the optimal solution consistently appearing among the top three most frequently sampled bitstrings, whereas the pure XY mixer shows much higher variability and worse ranking.

Although other studies have reported slightly better results for the standard XY mixer~\cite{qian2023comparative}, it is important to note that such works often employ a \textit{layerwise learning}, where the optimal parameters from depth $p$ are reused to initialize the optimization at depth $p+1$. In our case, each depth was optimized independently, without parameter reuse, to ensure a uniform comparison across methods. Despite this, our results remain very superior to those reported for the XY mixer, and are comparable to the RS-mixer for 5-city TSP instances and up to 3 layers. This comparison is particularly meaningful, as the RS-mixer is specifically constructed to allow transitions between all valid TSP solutions from any valid initial state, thereby enabling full exploration of the feasible solution space. However, this increased expressiveness comes at a significant computational cost. For a 16-qubit instance, each QAOA layer using the RS-mixer requires 1728 two-qubit gates, compared to just 64 two-qubit gates per layer when using the XY mixer \cite{qian2023comparative}. This substantial difference in circuit complexity makes the RS-mixer less suitable for near-term quantum hardware. The fact that our warm-started XY mixer approach, which is significantly more resource-efficient, achieves comparable results highlights its practical advantage. This makes the observed superiority of our combined approach even more notable and paves the way to potential benefits of hybridizing complementary QAOA strategies for solving structured combinatorial problems like the TSP.

An important question often raised in the context of warm-start techniques is whether QAOA can outperform the classical algorithm used to generate the warm-start in the first place. This issue was already addressed in the original warm-starting study, which demonstrated that, for the regularization factor $\epsilon = 0.25$, the same value used in our work, the quantum algorithm can indeed improve upon the classical approximation. This phenomenon is also clearly observed in our results. 

As shown in Table~\ref{tab:seeds_53_54_comparison}, even when initialized from suboptimal warm-starts, such as those generated for graph seeds 53 and 54, our approach still outperformed all other methods. Compared to the pure warm-start and pure XY mixer variants, the warm-started XY mixer achieved significantly lower ranks and higher sampling frequencies across all tested layers. These results not only confirm the robustness of our method under imperfect initializations, but also illustrate its clear advantage over each individual strategy when applied in isolation. The superior performance observed in these challenging instances further highlights the strength of combining warm-starting with a constraint-preserving mixer.

\begin{table}[ht]
\centering
\begin{threeparttable}
\caption{Performance comparison for seeds 53 and 54 across QAOA variants.}
\begin{tabular}{|c|c|c|c|c|}
\hline
\textbf{Seed} & \textbf{Layer} & \textbf{Approach} & \textbf{Rank} & \textbf{Frequence} \\
\hline
\multirow{3}{*}{53} & \multirow{3}{*}{1} & XY Warm-start & 11 & 17 \\
                    &                    & Pure XY       & 30 & 7  \\
                    &                    & Pure Warm     & 47 & 6  \\
\hline
\multirow{3}{*}{53} & \multirow{3}{*}{2} & XY Warm-start & 11 & 15 \\
                    &                    & Pure XY       & 26 & 9  \\
                    &                    & Pure Warm     & 352 & 6 \\
\hline
\multirow{3}{*}{53} & \multirow{3}{*}{3} & XY Warm-start & 5  & 31 \\
                    &                    & Pure XY       & 4  & 20 \\
                    &                    & Pure Warm     & 567 & 3 \\
\hline
\multirow{3}{*}{54} & \multirow{3}{*}{1} & XY Warm-start & 16 & 9  \\
                    &                    & Pure XY       & 35 & 7  \\
                    &                    & Pure Warm     & 3130 & 1 \\
\hline
\multirow{3}{*}{54} & \multirow{3}{*}{2} & XY Warm-start & 9  & 16 \\
                    &                    & Pure XY       & 60 & 6  \\
                    &                    & Pure Warm     & -- & -- \\
\hline
\multirow{3}{*}{54} & \multirow{3}{*}{3} & XY Warm-start & 14 & 12 \\
                    &                    & Pure XY       & 30 & 8  \\
                    &                    & Pure Warm     & 420 & 2 \\
\hline
\end{tabular}
\begin{tablenotes}
\item[*] Frequency values are based on 1000 samples per experiment.
\end{tablenotes}
\label{tab:seeds_53_54_comparison}
\end{threeparttable}
\end{table}

These specific instances also account for some of the outliers in the Sampling Rank results (Figure~\ref{fig:comparison_warm_vs_no_warm}, left), as well as for the larger standard deviations observed in the Percentage of True Solutions plot for the warm-started XY mixer (Figure~\ref{fig:comparison_warmXY_vs_XY}, right).

\section{Conclusion}
\label{sec:conclusion}

In this work, we have presented the {Warm-Starting QAOA with XY Mixers}, an hybrid quantum-classical approach that combines a warm-start initialization with a constraint-preserving XY mixer ansatz to improve the performance of QAOA on structured combinatorial problems. By constructing an initial quantum state that incorporates classical guidance from a Goemans–Williamson MaxCut relaxation and remains within the feasible subspace defined by one-hot TSP encodings, our method effectively balances solution quality and hardware efficiency.

Numerical simulations on 5-city TSP instances demonstrated that the proposed approach consistently outperforms both the standard XY mixer and the pure warm-starting variant in terms of solution quality and sampling accuracy. These results hold even under challenging conditions where the initial classical approximation is suboptimal, highlighting the robustness of the method.

Given its low circuit depth, limited gate overhead, and compatibility with constrained subspaces, the warm-started XY mixer strategy offers a promising path forward for NISQ-era quantum optimization. In future work, this approach could be extended to other constrained optimization problems, particularly those that already benefit from XY mixer-based ansätze and initial states constrained to fixed Hamming-weight subspaces. Such problems include variants of the Vehicle Routing Problem, scheduling tasks, and resource allocation models, where constraint-aware quantum evolution is crucial for generating feasible and high-quality solutions.

\appendix

\section{The MaxCut Problem}
\label{app:maxcut}
The MaxCut problem is a canonical NP-hard problem in combinatorial optimization with broad applications in physics, circuit design, and machine learning \cite{lucas2014ising, kochenberger2014qubo}. Given a weighted undirected graph $G = (V, E)$ with edge weights $\omega_{ij}$, the goal is to divide the set of nodes $V$ into two disjoint subsets such that the total weight of edges crossing the cut is maximized. Formally, this is expressed as the following binary quadratic optimization problem:
$$
\max_{z \in \{-1,1\}^{|V|}} \ \frac{1}{2} \sum_{(i,j) \in E} \omega_{ij}(1 - z_i z_j),
$$
where each binary variable $z_i$ indicates the subset assignment of vertex $i$. An edge contributes to the objective only when its endpoints are assigned to opposite partitions ($z_i \ne z_j$).

Despite its simple formulation, MaxCut remains computationally intractable for large graphs. Notably, even for graphs with positive edge weights, no classical polynomial-time algorithm can approximate MaxCut better than a $16/17$ ratio unless $\text{P} = \text{NP}$. This limitation has motivated a wide range of approximation strategies, including quantum approaches such as QAOA and warm-started quantum methods explored in this work.

\section{Converting QUBO to MaxCut}
\label{app:qubo_to_maxcut}
This appendix outlines the algebraic derivation that maps a generic QUBO problem into an equivalent instance of the MaxCut problem. Here we are following the derivation given in \cite{QUBO_to_Maxcut}.

\subsection{Step 1: Defining the Original QUBO Problem}
We start with the standard QUBO objective function:
\begin{equation}
f(x) = \sum_{j=1}^{n} \sum_{i=j+1}^{n} q_{ij} x_i x_j + \sum_{i=1}^{n} c_i x_i,
\end{equation}
where $x_i \in \{0,1\}$. Note that diagonal terms $q_{ii}$ are absorbed into the linear coefficients $c_i$.

\subsection{Step 2: Change of Variables}
To align the problem with MaxCut, we transform binary variables $x_i \in \{0,1\}$ to spin variables $s_i \in \{-1, +1\}$ using:
$$
s_i = 2x_i - 1 \quad \Longleftrightarrow \quad x_i = \frac{s_i + 1}{2}.
$$

Substituting into $f(x)$ gives:
\begin{multline}
f(s) = \sum_{j=1}^{n} \sum_{i=j+1}^{n} \left[ \frac{1}{4} q_{ij} s_i s_j 
+ \frac{1}{4} q_{ij} s_i + \frac{1}{4} q_{ij} s_j + \frac{1}{4} q_{ij} \right] \\
+ \sum_{i=1}^{n} \left( \frac{1}{2} c_i s_i + \frac{1}{2} c_i \right)
\end{multline}

Expanding terms yields:
\begin{multline}
f(s) = \sum_{j=1}^{n} \sum_{i=j+1}^{n} \frac{1}{4} q_{ij} s_i s_j 
+ \sum_{i=1}^{n} \left[ \frac{1}{4} \left( \sum_{j \ne i} q_{ij} \right) 
+ \frac{1}{2} c_i \right] s_i \\
+ \left[ \sum_{j=1}^{n} \sum_{i=j+1}^{n} \frac{1}{4} q_{ij} 
+ \sum_{i=1}^{n} \frac{1}{2} c_i \right]
\end{multline}

\subsection{Step 3: Grouping Terms}
Rewriting $f(s)$ to isolate terms involving $s_i s_j$, $s_i$, and constants:
\begin{align}
f(s) = &\sum_{j=1}^{n} \sum_{i=j+1}^{n} \frac{1}{4} q_{ij} s_i s_j + \sum_{i=1}^{n} \left[ \frac{1}{4} \left( \sum_{j \ne i} q_{ij} \right) + \frac{1}{2} c_i \right] s_i \nonumber\\
&+ \left[ \sum_{j=1}^{n} \sum_{i=j+1}^{n} \frac{1}{4} q_{ij} + \sum_{i=1}^{n} \frac{1}{2} c_i \right].
\end{align}

Define the constant term $C_1$ and new linear coefficients $d_i$ as:
$$
C_1 = \sum_{j=1}^{n} \sum_{i=j+1}^{n} \frac{1}{4} q_{ij} + \sum_{i=1}^{n} \frac{1}{2} c_i, \quad
d_i = \frac{1}{4} \sum_{j \ne i} q_{ij} + \frac{1}{2} c_i.
$$

\subsection{Step 4: Introducing Auxiliary Node $s_0$}
To cast the objective in MaxCut form, we add an auxiliary variable $s_0 \in \{-1, +1\}$, fixed to $s_0 = +1$. Define:
$$
g(s) = \sum_{j=1}^{n} \sum_{i=0}^{j-1} w_{ij} s_i s_j + C_1,
$$
with new weights $w_{ij}$ defined as:
$$
w_{ij} = \frac{1}{4} q_{ij}, \quad (1 \leq i < j \leq n),
$$
$$
w_{0j} = \frac{1}{4} \left( \sum_{i=1}^{j-1} q_{ij} + \sum_{i=j+1}^{n} q_{ji} \right) + \frac{1}{2} c_j, \quad (1 \leq j \leq n).
$$

\subsection{Step 5: Mapping to MaxCut}
We now identify the objective $g(s)$ with the standard MaxCut form. Define the cut as the set of edges between two partitions:
$$
V^+ = \{ i \mid s_i = +1 \}, \quad V^- = \{ i \mid s_i = -1 \}.
$$
Assuming $s_0 = +1$, we fix node 0 in $V^+$ without loss of generality.

The cut value is then:
$$
c(\delta(W)) = \sum_{i \in V^+} \sum_{j \in V^-} w_{ij}.
$$
Thus, minimizing $f(x)$ is equivalent to maximizing the cut value $c(\delta(W))$.

Finally, we recover the original QUBO variables from the MaxCut partition via:
$$
x_i = \begin{cases}
1, & \text{if } i \in V^-,\\
0, & \text{if } i \in V^+.
\end{cases}
$$

\bibliography{IEEEexample}

\end{document}